\documentclass[aps, showpacs, preprint,preprintnumbers,amsmath,amssymb,floatfix]{revtex4}
\usepackage{graphicx}
\usepackage{wrapfig}
\usepackage{epsfig}
\usepackage{dcolumn}
\usepackage{tabls}
\usepackage[section]{placeins}
\newcommand{\be}{\begin{eqnarray}}
\newcommand{\ee}{\end{eqnarray}}
\begin{document}
\title{\bf Ab initio results for the broken phase of scalar
light front field theory}
\author{Dipankar Chakrabarti}
\email{dipankar@phys.ufl.edu}
\affiliation{Physics Department, University of Florida, Gainesville, FL
32611, U.S.A. }
\author{A. Harindranath}
\email{hari@theory.saha.ernet.in}
\affiliation{Theory Group, Saha Institute of Nuclear Physics,
1/AF Bidhan Nagar, Kolkata 700064, India}

\author{ L$\!\!$'ubomir Martinovi\v c}
\email{fyziluma@savba.sk}
\affiliation{Institute of Physics, Slovak Academy of Sciences,
D\'ubravsk\'a cesta 9, 845 11 Bratislava, Slovakia}

\author{Grigorii B. Pivovarov}
\email{gbpivo@ms2.inr.ac.ru}
\affiliation{Institute for Nuclear Research, Moscow 117312 Russia}

\author{James P. Vary}
\email{jvary@iastate.edu}
\affiliation{Department of Physics and Astronomy, Iowa State University,
Ames, IA 50011, U.S.A.}

\date{May 5, 2005}

\begin{abstract}
We present nonperturbative light-front energy eigenstates
in the broken phase of a two dimensional $\frac{\lambda}{4!}\phi^4$
quantum
field theory using Discrete Light Cone Quantization and
extrapolate the
results to the continuum limit.
We establish degeneracy in the even and odd particle sectors and extract
the masses of the lowest two states and the vacuum energy density for
$\lambda=0.5$ and 1.0. We present
two novel results: the Fourier transform of the form factor of the
lowest excitation as well as the number density of elementary
constituents of
that state. A coherent state with kink - antikink structure is revealed.
\end{abstract}
\maketitle
\vskip .1in

Quantum descriptions of spontaneous symmetry breaking (SSB) and
topological objects play a major role in condensed
matter physics, quantum field theory and string theory. Since a direct
solution of the quantum problem is extremely difficult, traditional
approaches rely on semi-classical expansions, variational
approximations like the coherent state ansatz,
self-consistent methods like the Hartree approximation, etc. Using these
methods, one can test for the presence of SSB
and compute the leading quantum corrections to topological objects
in a model. One would like to know the reliability of these
approximation
schemes, to calculate the masses nonperturbatively and also to determine
additional observables from Hamiltonian eigenfunctions.
For example, with the eigenfunctions, one could study whether the
number density of $``$elementary
constituents" follows the behavior suggested by the
variational coherent state ansatz.

We will use the Discrete Light Cone Quantization method (DLCQ)
\cite{dlcq}
to address these issues. This method can provide exact (nonperturbative)
predictions, in the continuum limit, of the masses of the
lowest states and vacuum energy density. 

In the light front literature it has been suggested that (see the 
review in Ref. \cite{dlcq}) without a field
mode carrying exactly zero momentum, it would be impossible to 
describe spontaneous symmetry breaking. Rozowsky and Thorn \cite{RT} pointed 
out that if such a mode were really necessary, one of the attractive 
features of light front formulations of field theory and string theory, 
namely, casting any relativistic quantum mechanical system in a 
Newtonian picture, would no longer be viable. They convincingly
argued that the 
physics of condensation does not require a mode with exactly zero momentum.
They performed an analytic variational calculation at weak
coupling to corroborate their hypothesis and we utilize their
coherent state ansatz to help interpret our results.
However, the numerical calculations carried out by these authors to
independently 
verify this hypothesis were not conclusive since the volumes 
considered were not large enough to overcome tunneling.

One of our main results is that
we do observe SSB even without the explicit presence of the notorious
zero-momentum mode as anticipated by Rozowsky and Thorn \cite{RT}.
We also compute the Fourier transform of the form factor
of the lowest state (i.e., the $``$ profile" of the kink-antikink
configuration) and its parton content, results not yet available from
other methods. In addition, at weak coupling, we extract the value of the 
condensate from two observables, (1) the computed vacuum energy density, 
and (2) the asymptote of the profile function.

As is well known \cite{dashen,gj}, topological excitations (kinks) exist
in the
classical as well as quantum two-dimensional $\phi^4$ model
with negative quadratic term (broken phase). It was proven rigorously
that in quantum theory a stable kink state is
separated from the vacuum by a mass gap of the order $\lambda^{-1}$ and
from the rest of the spectrum by an upper gap \cite{Froh}. More detailed
nonperturbative information on the spectrum of the mass operator or on
other
observables from rigorous approaches is not available. 

A popular nonperturbative numerical approach to field theory is 
the Euclidean lattice formulation. In the topologically non-trivial 
sector of two-dimensional
$\phi^4$ theory, results available from lattice simulations, so far 
\cite{lattice}, are limited to the determination of the kink mass.  The results 
for the configuration average of the kink profile are not smooth and 
are difficult to interpret, perhaps due to problems with 
thermalization and/or finite volume limitations.

      These Euclidean lattice calculations are highly non-trivial and 
a brief overview displays the degree of effort needed to reveal 
topological observables. In one approach, one computes the kink mass 
from the decay of the correlation functions of an operator with 
nonvanishing projection on the topological sector under 
consideration, the dual field in the present case. On a finite 
lattice, the definition of such an operator is often ambiguous. 
Another approach involves integrating the difference between the 
expectation values of the lattice actions with antiperiodic and 
periodic boundary conditions. To obtain results for the continuum 
field theory, one has to work in the critical region of the lattice 
theory. Here, calculations are severely hampered by the phenomena of 
critical slowing down. Given these difficulties, it is understandable 
that Euclidean lattice calculations of the mass and other properties 
of the kink-antikink state have not been reported to date.
   The strengths of the Euclidean lattice approach lie in extracting 
critical properties and low mass eigenstates of the theory.  On the 
other hand, the strengths of DLCQ are complementary and lie in 
extracting a range of observables for the mass eigenstates.

In the variational approach, kinks can be well-approximated by coherent
states. This appears to have two implications for the Fock space
expansion in
our discretized approach, namely,
(a) one may need an infinite number of bosons to describe
solitons, and, (b) since the dimensionless total longitudinal momentum
$K$
automatically provides a cutoff on the number of bosons, convergence in
$K$
may be difficult to achieve for a kink-like state \cite{Burkardt}.
Here, we show how a
nonperturbative evaluation of topological excitations
and their observables is feasible in a finite Fock basis.

We investigate these issues in 
two dimensional $\phi^4$ theory using DLCQ
with periodic boundary condition (PBC). We adopt the convention
$ x^\pm = x^0 \pm x^1$. The Lagrangian density is
\be {\cal L} = \frac{1}{2} \partial^\mu \phi \partial_\mu \phi +
\frac{1}{2}
\mu^2 \phi^2 - \frac{\lambda}{4!} \phi^4.
\ee
The Hamiltonian density $ {\cal P}^- = - (1/2)
\mu^2 \phi^2  + ({\lambda}/{4!} )\phi^4 $ and the momentum density
$ {\cal P}^+ = \partial^+ \phi \partial^+ \phi$. In DLCQ, the Hamiltonian
operator with dimension of $mass^2$, $ H = (2 \pi /L) P^ - = (2 \pi /L)
\int_{-L}^{+L} dx^- {\cal P}^-$ and the dimensionless momentum operator
$ K = (L/2 \pi) P^+ = (L/2 \pi) \int _{-L}^{+L} dx^- {\cal P}^+$. Here
$L$
denotes our compact domain $ - L \le x^- \le +L$. Throughout this work we
address the eigen solutions of $H$ with $ \mu^2=1$.

The scalar field can be decomposed as $\phi(x^-) = \phi_0 +
\Phi(x^-)$, where $\phi_0$ is the zero mode operator omitted in our
Hamiltonian and $\Phi(x^-)$ is the normal mode operator
\be
\Phi(x^-) = \frac{1}{\sqrt{4 \pi}} \sum_{n} \frac{1}{\sqrt{n}}
\left [a_n e^{-i \frac{n \pi}{L} x^-} + a_n^\dagger e^{i \frac{n
\pi}{L} x^-} \right ].
\ee

The normal ordered Hamiltonian  is given by 

\be
H & = & - \mu^2 \sum_n \frac{1}{n} a_n^\dagger a_n
+ \frac{\lambda}{4 \pi} \sum_{k \le l, m\le n}~ \frac{1}{N_{kl}^2}
~ \frac{1}{N_{mn}^2}~
\frac{1}{\sqrt{klmn}}
 a_k^\dagger a_l^\dagger a_n a_m
~\delta_{k+l, m+n} \nonumber \\&~& + \frac{\lambda}{4 \pi} \sum_{k, l \le
m\le n}~ \frac{1}{N_{lmn}^2}~
\frac{1}{\sqrt{klmn}}~
  \left [
a_k^\dagger a_l a_m a_n + a_n^\dagger a_m^\dagger a_l^\dagger a_k \right ]~
\delta_{k, l+m+n}
\ee
with
\be N_{lmn} & = & 1 ,~ l \ne m \ne n, \nonumber \\
            & = & \sqrt{2!},~ l=m \ne n, ~ l \ne m=n,\nonumber \\
            & = & \sqrt{3!},~ l=m=n,
\ee
and
\be
N_{kl} & = & 1, k \ne l,~ \nonumber \\
       & = & \sqrt{2!},~ k=l.
\ee


We diagonalize this Hamiltonian in the basis of all many-boson
configurations at a fixed $K$ where $K$ is the sum of the values 
of the dimensionless momenta of
all bosons in the configuration. 
The Hamiltonian is symmetric under $ \phi \rightarrow -\phi$ and thus, 
with PBC, the
Hamiltonian matrix becomes block diagonal in even and odd particle number
sectors. 
\tablinesep=.1in
\arraylinesep=.1in
\extrarulesep=.1in
\begin{table}
\begin{tabular}{||c|c||c|c||}
\hline \hline
\multicolumn{2}{||c||}{even sector} & \multicolumn{2}{c||}{odd sector}
\\
\hline
 K & dimension & K & dimension \\
\hline
 16 & 118 & 15 & 90 \\
 24 & 793 & 23 & 632 \\
 32 & 4186 & 31 & 3431\\
 40 & 18692 & 39 & 15613\\
 50 & 102162 & 49 & 86809\\
 60 & 483338 & 59 & 416006 \\
 64 & 870953 & 64 & 870677 \\
 68 & 1544048 & 68 & 1543687 \\
\hline\hline
\end{tabular}
\caption{Dimensionality of the Hamiltonian matrix in even and odd
particle
sectors with periodic boundary condition.}
\label{dim}
\end{table}
The dimensionality of the matrix in the
even and odd sectors for different $K$ is presented in Table \ref{dim}.
Data in Table \ref{dim} agrees with the earliest work \cite{hv} on two
dimensional $\phi^4$ theory in DLCQ which presented results for $ K (\le
16)$.  
We obtained our results on clusters of computers
($\le 28$ processors) using the Many Fermion Dynamics (MFD) code adapted
to bosons \cite{mfd}. The Lanczos method is used in a highly scalable
algorithm.

Since we dropped the $P^+=0$ mode, degenerate vacuum states,
characterized
by a spatially uniform field expectation value, are not explicitly
present
in our formulation. However, we expect degeneracy of the energy levels in
the
even and odd particle sectors at sufficiently high resolution, $K$.
The argument is as follows. For small coupling,
variational coherent states $\vert\alpha\rangle$ represent a good
approximation of the lowest lying physical states \cite{RT}. Even and odd
states are linear combinations of $\vert \alpha \rangle$ and $\vert
-\alpha
\rangle$ and for large enough $K$ they have the same energy.
We also expect that our lowest state
will be an excitation above the vacuum state and we will
show it corresponds to a configuration with properties of a
kink-antikink pair.
\begin{figure}
\centering
\includegraphics[width=3.8in,clip]{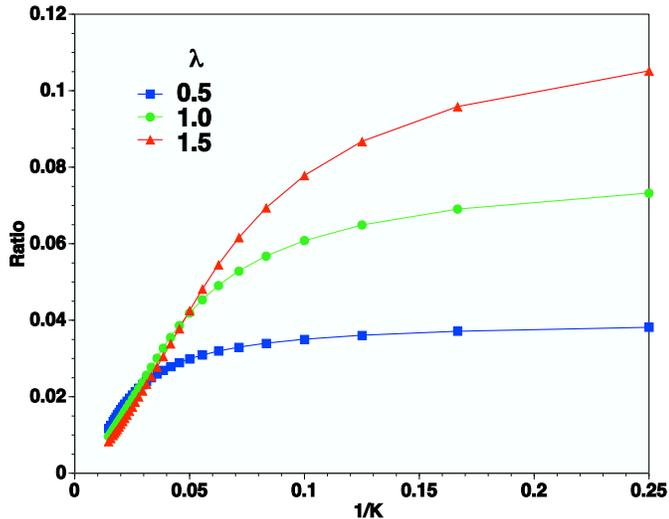}
 \caption{ Ratio of lowest state 
even-odd energy difference to classical
energy density for
$\lambda=0.5, 1.0, 1.5$ .}
\label{ratio}  
\end{figure}  
In Fig. \ref{ratio} we present a ratio, the difference between the lowest
eigenvalues of different parity divided by the classical vacuum energy
density, as a function of the inverse resolution. Curves for $\lambda =
0.5, 1.0, 1.5$ all demonstrate the trend to degeneracy in the continuum
limit
($K \rightarrow \infty $).
That is, we obtain SSB at each coupling through degeneracy of even and
odd parity states when we extrapolate to the continuum limit.
At any finite $K$ the lifting of the degeneracy is simply a reflection of
the tunneling present in a finite system. As seen from Fig. \ref{ratio}, 
the tunneling is relatively strong for $K \le 20$.

\begin{figure}
\centering
\epsfig{figure=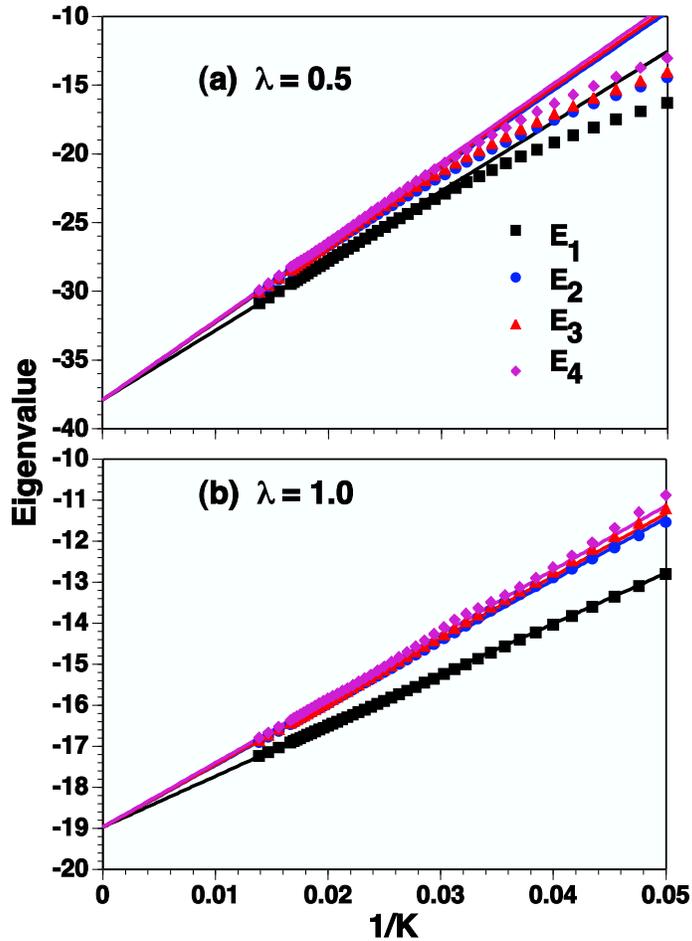,width=3.5in}
 \caption{ Lowest four eigenvalues versus $\frac{1}{K}$
for
(a) $\lambda=0.5$. (b)
$\lambda=1.0$.}
\label{4evs}
\end{figure}

Now consider the behavior of the lowest four eigenvalues with $K$ for
$\lambda=0.5,1.0$ as presented in Fig. \ref{4evs}.
For 20 $ \le K \le 72$ we display the even particle number results at
even
$K$ (which are lower than the odd particle results at the same $K$) and
the odd particle number results at odd $K$. The results follow smooth
curves becoming more linear as $K$ increases as seen by the straight
lines from
fitting the $ 50 \le K \le 72 $ results.

Our lowest state is expected to be a kink-antikink configuration and
should have a positive invariant mass (twice the mass of the single
kink at weak coupling).
For massive states, the light front energy $E$ scales like
$(1/K)$ so that it approaches zero in the infinite $K$ limit. On the
other
hand, a coherent
state variational calculation shows that in the infinite $K$ limit, the
energy of the lowest state approaches the classical ground state
energy density $ {\cal E}=
- (6 \pi \mu^4 / \lambda)$ for small $\lambda$.
As we show in what follows, our results are increasingly compatible with
a
coherent state as $K$ increases towards the continuum limit.
Thus, we fit our finite $K$ results for the eigenvalues at small
$\lambda$
to the formula $C + M^2/K$, where $C$ is the vacuum energy density
and $M$ is the kink-antikink mass.
\tablinesep=.1in
\arraylinesep=.1in
\extrarulesep=.1in
\begin{table}[bht]
\centering
\begin{tabular}{||c|c|c|c|c|c||}
\hline \hline
 $\lambda$ & \multicolumn{2}{c|}{vacuum energy} &
\multicolumn{3}{c||}{soliton mass}\\
\hline
& classical & this work & classical & semi-class. & this work\\
\hline
0.5 & -37.70 & -37.90(4) & 11.31 & 10.84 & 11.26(4) \\
1.0 & -18.85 & -18.97(2) & 5.657 & 5.186 & 5.563(7) \\
\hline
\hline
\end{tabular}
\caption{Comparison of vacuum energy and soliton mass from
the continuum limit of our results, with classical results.
Semi-classical
results for the mass \cite{dashen} are also shown. Our estimated
uncertainties in the last significant digit are quoted in parenthesis. }
\label{tab2}
\end{table}
Extracted values of $C$ and the kink or soliton mass
$M/2$ are compared to their classical and one loop corrected
($``$semi-class.") counterparts in Table \ref{tab2}.
Uncertainties, quoted in parenthesis,
are estimated from experience after
performing a variety of extrapolations.
Our vacuum energy nearly coincides
with the classical result probably as the result of dropping the zero
mode.
$M/2$ is
also close to the classical value since our coupling constant is
small and the kink-antikink interactions are weak.
Our soliton mass and vacuum energy 
at $\lambda =1$  are in reasonable
agreement with  results using antiperiodic boundary conditions
\cite{apbc}.
The fact that the mass of the quantum kink is larger than the
semi-classical
one is peculiar to the choice
$\mu^2=1$
and does not occur for
$\mu^2$ away from  $1$ \cite{lattice}.
Using similar fits to the next excited state,
constrained to have the same
$ 1/K \rightarrow 0$ intercept (vacuum energy), we obtain the $M^2$
gap. For $\lambda$=0.5 (1.0) it is 57.3 (26.6). These
results are approximately 89\% (83\%) of the corresponding variational
estimates of Ref. \cite{RT}. 

\begin{figure}
\centering
\includegraphics[width=3.5in, clip]{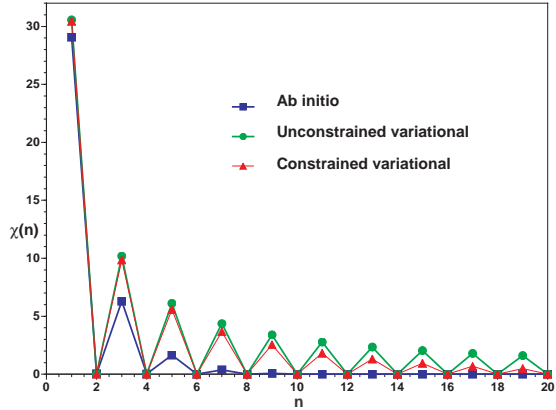}
 \caption{ Comparison of the number density $\chi(n)$
from
our approach ("Ab initio") ($ K=60$)
and the constrained and unconstrained coherent state variational
calculation for $ \lambda=1.0.$   }
\label{cohcomp}
\end{figure}

As an example of another observable, we evaluate the occupation
number density $\chi(n)$, the 
analog of the parton distribution
function of more realistic theories.
Note that in the unconstrained variational  
state \cite{RT}, the shape of  $\chi(n)$ is independent of the
coupling $\lambda$ which affects only its overall normalization. On the
other hand, in the variational calculation, constrained to have a fixed
value of $ \langle K \rangle$, $\lambda$ affects not only
the overall normalization but also the shape of the distribution.
In Fig. \ref{cohcomp} our result at $K=60$
is compared with that of the unconstrained
and constrained ($ \langle K \rangle =60$) coherent
state approximation for $ \lambda=1.0.$
We find that the shape and normalization of our $\chi(n)$ depends on
$\lambda$. Our results display the same sawtooth
pattern as the constrained and 
unconstrained variational results.
This is due to the PBC and  
the sawtooth pattern is not present in the number density in the
case of antiperiodic BC \cite{apbc}. We do expect sensitivity to the
boundary conditions in topologically  non-trivial sectors
even in the infinite volume limit.

Another observable that yields information about the spatial
structure of the low lying states
is the Fourier transform of the form
factor.
We compute this observable for the lowest state which,
 according to Goldstone and Jackiw
\cite{gj},
represents the classical kink-antikink profile
in the weak coupling 
limit. Let $ \mid K \rangle$ and $ \mid K' \rangle$ denote this state
with momenta $K$ and $K'$.
In the continuum, 
\begin{eqnarray}
\int_{- \infty}^{+ \infty} dq^+ exp\{- \frac{i}{2} q^+ a\}
\langle K \mid \Phi(x^-) \mid K' \rangle = \phi_c(x^- - a). \nonumber
\end{eqnarray}
For a study of the form factor of topological objects 
in the semi-classical approximation at
finite volume with relevance to the work of Goldstone and Jackiw, see Ref.
\cite{mrs}.

\begin{figure}
\centering
\includegraphics[width=3.7in,clip]{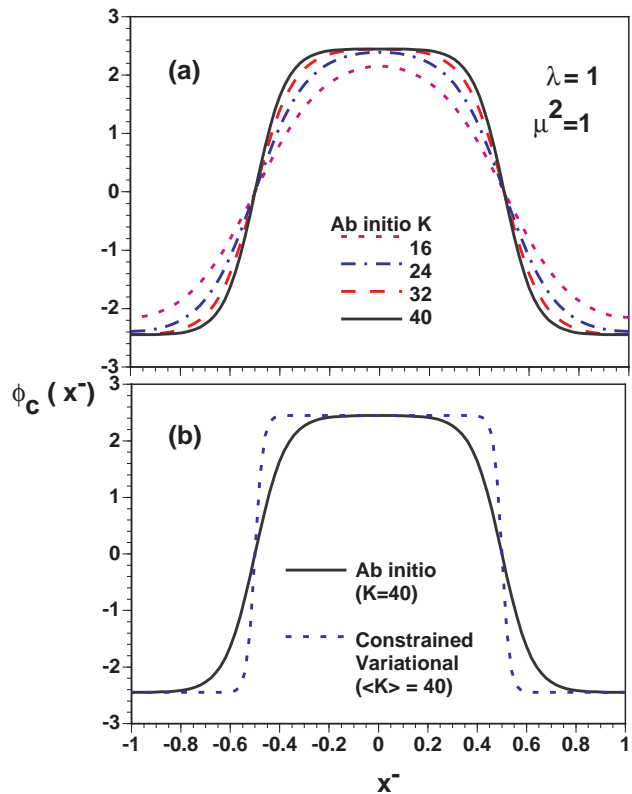}
 \caption{ Fourier Transform of the kink-antikink form
factor at
$\lambda=1.0$. Results are plotted in units of $L$. (a) Convergence with
$K$. (b) Comparison of our result (Ab initio)
($K$=40) with
constrained variational calculation ($ \langle K \rangle = 40$).   }
\label{profile}
\end{figure}

For the form factor, we require the same state at different $K$
values since $
K' = K+q$. We
proceed as follows. We diagonalize the Hamiltonian, say, at $K=40$ in the
even particle sector. Then we diagonalize the
Hamiltonian at the neighboring odd $K$ values, $K=31, 33, \dots 49$
in the odd particle sector so that the dimensionless
momentum transfer ranges from -9 to 9.
In the sum replacing the above integral in $q^+$, we employ states that
fall on the same linear trajectory in Fig. \ref{4evs} so that we can be
confident that all these states correspond to the same physical state.
We then compute the matrix element of the field
operator between the lowest state at $K=40$ and all other values of K
and sum the amplitudes with the shift parameter $a=0$.
Here, we need to be careful about the phases.
First we note that $K$ is a conserved quantity, so eigenfunctions at
different $K$ have an overall arbitrary complex phase factor.
To set the relative phases between pairs of eigenstates at different $K$,
we set the signs of the leading amplitude to be the same as
dictated by the coherent state variational calculation.
The topology of a kink - antikink structure and
other properties of the form factor rely on the detailed
behavior of the
amplitudes over a range of $K$ values.
The result for the lowest eigenstate for
$\lambda=1$ is presented in Fig. \ref{profile}. This striking kink - antikink
behavior is particular to the lowest state.

Taking the vacuum energy results of Table \ref{tab2} together with the
$\phi_c(x^-)$
results of Fig. \ref{profile}, we have two independent but consistent
methods for extracting $ \langle \phi \rangle$ in the weak coupling
limit.
From our vacuum energy density for $\lambda=1.0$, we obtain $
\langle \phi \rangle = ({\cal E}/ \pi \mu^2)^{1/2} = 2.457$. From the
calculation of the profile function, shown in Fig. \ref{profile}, we
extract
$ \langle \phi \rangle $ as the asymptotic ($ x^- = \pm 1 $ in units of
$L$)
intercepts to be equal to 2.447. These results may be compared with the
classical value of $ \sqrt{6}=2.449$, which agrees with the result from
the variational coherent state.

In summary, we have demonstrated the phenomenon of spontaneous symmetry
breaking in a discretized light front approach without $P^+$
zero mode and calculated several nonperturbative physical quantities.
We find that a finite Fock space yields
features of the lowest excitation that are similar
to those of a variational coherent state
ansatz. We have extracted the quantum kink mass and the vacuum energy
density for
small $\lambda$ by extrapolating our lowest eigenvalue to the continuum
limit.
At weak coupling, the mass of the quantum kink is closer to the
classical value than to the semi-classical mass. We have extracted the number
density of elementary constituents of the lowest state and compared it
with the coherent state prediction. We have also evaluated the Fourier
transform of the lowest state form factor in a fully non-perturbative
quantum approach and obtained a kink-antikink profile.

This work is supported in part by the Indo-US
Collaboration project jointly funded by the U.S. National Science
Foundation (NSF) (INT0137066) and the Department of Science and
Technology,
India
(DST/INT/US (NSF-RP075)/2001). This work is also supported in part by the
US
Department of Energy, Grant No. DE-FG02-87ER40371 and the U.S. NSF Grant
No. NSF PHY 007 1027.
Two of the authors (D.C. and A.H.) would like to acknowledge many useful
discussions with Asit K. De.
The work of G.P. was supported by Russian Foundation for Basic Research,
project No.03-02-17047. L.M. was partially supported by the
APVT grant No. 51-005704.


\end{document}